# Time and Classical and Quantum Mechanics: Indeterminacy vs. Discontinuity

**Peter Lynds**[1]

Time, Classical Mechanics, Quantum Mechanics, Indeterminacy, Discontinuity, Relativity, Cosmology, Imaginary Time, Chronons, Zeno's Paradoxes.

It is postulated there is not a precise static instant in time underlying a dynamical physical process at which the relative position of a body in relative motion or a specific physical magnitude would theoretically be precisely determined. It is concluded it is exactly because of this that time (relative interval as indicated by a clock) and the continuity of a physical process is possible, with there being a necessary trade off of all precisely determined physical values at a time, for their continuity through time. This explanation is also shown to be the correct solution to the motion and infinity paradoxes, excluding the Stadium, originally conceived by the ancient Greek mathematician Zeno of Elea. Quantum Cosmology, Imaginary Time and Chronons are also then discussed, with the latter two appearing to be superseded on a theoretical basis.

## 1. Introduction

Time enters mechanics as a measure of interval, relative to the clock completing the measurement. Conversely, although it is generally not realized, in *all* cases a time value indicates an interval of time, rather than a precise static instant in time at which the relative position of a body in relative motion or a specific physical magnitude would theoretically be precisely determined. For example, if two separate events are measured to take place at either 1 hour or 10.00 seconds, these two values indicate the events occurred during the time intervals of 1 and 1.99999…hours and 10.00 and 10.0099999…seconds, respectively. If a time measurement is made smaller and more accurate, the value comes closer to an accurate measure of an interval in time and the corresponding parameter and boundary of a specific physical magnitudes potential measurement during that interval, whether it be relative position, momentum, energy or other. Regardless of how small and accurate the value is made however, it cannot indicate a precise static instant in time at which a value would theoretically be precisely determined, because there is not a precise static instant in time underlying a dynamical physical process. If there were, all physical continuity, including motion and variation in all physical magnitudes would not be possible, as they would be frozen static at that precise instant, remaining that way. Subsequently, at no time is the relative position of a body in relative motion or a physical magnitude precisely determined, whether during a measured time interval, however small, or at a precise static instant in time, as at no time is it not constantly changing and undetermined. Thus, it is exactly due to there *not* being a precise static instant in time underlying a dynamical physical process, and the relative motion of body in relative motion or a physical magnitude *not* being precisely determined at any time, that motion and variation in physical magnitudes is possible: there is a necessary trade off of all precisely determined physical values at a time, for their continuity through time.

In the present report this simple but very counter-intuitive conclusion is developed and explored in further detail and its general implications have important significance to time and its relationship to classical and quantum mechanics, while also providing an insight into the reason and purpose for indeterminacy and uncertainty in nature. An overview of the main theoretical results reported, presented in the numerical order in which they later appear follows: (3) A body (micro and macroscopic) in relative motion does not have a precisely determined relative position at any time, and all physical magnitudes are not precisely determined at any time, although with the parameter and boundary of their respective position and magnitude being determinable up to the limits of possible measurement as stated by the general quantum hypothesis and Heisenberg's uncertainty principle[1], but with this indeterminacy in precise value not being a consequence of $\hbar$ and quantum uncertainty. This illustrates that in relation to indeterminacy in *precise* physical magnitude, the micro and macroscopic are inextricably linked, both being a part of the same parcel, rather than just a case of the former underlying and contributing to the latter. (3.1) The explanation provided is then also shown to be the correct solution to the motion and infinity paradoxes, excluding the Stadium, originally conceived by the ancient Greek mathematician Zeno of Elea. (4) It is not necessary for time to "emerge" and "congeal" out of the "quantum foam" and highly

[1] c/- 21 Oak Avenue, Paremata, Wellington, New Zealand. PeterLynds@xtra.co.nz : http://www.peterlynds.net.nz

contorted space-time geometry's present preceding Planck scale $(Gh/c^3)^{1/2}$ just after the big bang (new inflationary model), as has often previously been tentatively hypothesized.[2-7] Continuity would be present and naturally inherent in practically all initial quantum states and configurations, rather than a specific few, or special one, regardless of how microscopic the scale. (4.1) Furthermore, the cosmological proposal of "Imaginary Time",[2, 3, 5-7] is not compatible with a consistent physical description, both, as a consequence of the above consideration, and secondly, because it is the relative order of events that is relevant, not the direction of time itself. As a consequence, it is not possible for the order of a sequence of events to be imaginary (at right angles) relative to another sequence of events. (5) Lastly, "Chronons", proposed particles of indivisible intervals of time,[2, 8] also appear to be superseded on a theoretical basis, as their possible existence is incompatible with the simple conclusion that the very reason physical continuity is possible in the first instance is due to there not being a quantum or atom of time.

Before proceeding further however, I think it is important to stress that although I have attempted to be as quantitative and rigorous as possible, the subject of time does not readily lend itself to such a description, particularly in the context in which it is treated here, and readers may initially find that they will need to really grapple with the contents before they are able to achieve a clear and genuine understanding. I apologize for this, but I can find no other way of conveying the same information and laying the initial foundations for the physics that subsequently follow, of which, in relation to importance, I consider to quite easily outweigh any undesirable, although unavoidable and necessary aspects of this paper.

## 2. Motion and Continuity

We begin by considering the simple and innocuous postulate: 'there is not a precise static instant in time underlying a dynamical physical process.' If there were, the relative position of a body in relative motion or a specific physical magnitude, although precisely determined at such a precise static instant, would also by way of logical necessity be frozen static at that precise static instant. Furthermore, events and all physical magnitudes would remain frozen static, as such a precise static instant in time would remain frozen static at the same precise static instant. (Incidentally, the same outcome would also result if such a precise static instant were hypothetically followed by a continuous sequence of further precise static instants in time, as by their very nature a precise static instant in time does not have duration over interval in time, so neither could a further succession of them. This scenario is not plausible however in the first instance, as the notion of a *continuous progression* of precise static instants in time is obviously not possible for the same reason). Rather than facilitating motion and physical continuity, this would perpetuate a constant precise static instant in time, and as is the very nature of this ethereal notion i.e. a physical process frozen static at an 'instant', as though stuck on pause or freeze frame on a motion screen, physical continuity is not possible if such a discontinuous chronological feature is an intrinsic and inherent property of a dynamical physical process, and as such, a meaningful (and actual physical) indicator of a time at which the relative position of a body in relative motion or a certain physical magnitude is precisely determined, as has historically been assumed. That is, it is the human observer who subjectively projects, imposes and assigns a precise instant in time upon a physical process, for example, in order to gain a meaningful subjective picture or 'mental snapshot' of the relative position of a body in relative motion.

As a natural consequence of this, if there is not a precise static instant in time underlying a dynamical physical process, there is no physical progression or flow of time, as without a continuous and chronological progression through definite indivisible instants of time over an extended interval in time, there can be no progression. This may seem somewhat counter-intuitive, but it is exactly what is required by nature to enable time (relative interval as indicated by a clock), motion and the continuity of a physical process to be possible. Intuition also seems to suggest that if there were not a physical progression of time, the entire universe would be frozen motionless at an instant, again as though stuck on pause on a motion screen. But if the universe were frozen static at such a static instant, this would be a precise static instant of time: time *would* be a physical quantity. Thus, it is then due to natures very exclusion of a time as a fundamental physical quantity, that time as it is measured in physics (relative interval), and as such, motion and physical continuity are indeed possible.

It might also be argued in a more philosophical sense that a general definition of static would entitle a certain physical magnitude as being *unchanging* for an extended interval of time. But if this is so, how then could time itself be said to be frozen static at a precise instant if to do so also demands it must be unchanging for an extended interval of time? As a general and sensible definition this is no doubt correct, as we live in a world where indeed there is interval in time, and so for a certain physical magnitude to be static and unchanging it would naturally also have to remain so for an extended duration, however short. There is something of a paradox here however. If there were a precise static instant underlying a dynamical physical process, everything, including clocks and watches would also be frozen static and discontinuous, and as such, interval in time would not be possible either. There could be no interval in time for a certain physical magnitude to remain unchanging. Thus this general definition of static breaks down when the notion of static is applied to time itself. We are so then forced to search for a revised definition of static for this special temporal case. This is done by qualifying the use of stasis in this particular circumstance by noting static and unchanging, with static and unchanging as not being over interval, as there could be no interval and nothing could change in the first instance. At the same time however, it should also be enough just to be able to recognize and acknowledge the fault and paradox in the definition when applied to time.

It might also be argued by analogy with the claim by some people that the so-called 'block universe model', i.e. a 4-dimensional model of physical reality, incorporating time as well as space, is static or unchanging. This claim however involves the common mistake of failing to recognize that unless there is *another* time dimension, it simply doesn't make sense to say that the block universe is static, for there is no 'external' time interval over which it remains the same. If we then apply the same line of reasoning to the hypothetical case being discussed presently, we could say: It doesn't make sense to say that everything would be static at an instant, (with physical continuity and interval in time not being possible), as there would be no time interval for such an assertion to be relative to, referenced from, or over which such an instant would remain the same etc. This objection is valid. However, as it applies to the hypothetical case under investigation, it should also be clear that it is not any more applicable or relevant than being a semantical problem of the words one employs to best try to put across a point and as being a contradiction in terms, rather than pertaining to any contradiction in the actual (in this case, hypothetical) physics involved. One could certainly also assert that there were no interval in time, and so if one wishes, there were a precise static instant underlying a physical process, without it being dependent on there actually being interval: as is the case with the hypothetical absence of mass and energy, and the resulting absence of 3 spatial dimensions.

It is also important to note this conclusion is compatible with the dynamical manner in which time enters the equations, geometry and description of the universe in Albert Einstein's theories, special and general relativity.[9] It is relative interval as measured by all clocks (whether digital, atomic, light, biological or other) that is warped and mutable at relativistic velocities and in the spatial vicinity of gravity, not any physical progression of time. Indeed, it could be said it is due to there not being a physical instant and physical progression of time, that the continuity, propagation and constant relative velocity of electromagnetic radiation, and thus, a warping of relative duration is possible in the first instance. Subsequently, this conclusion is also consistent and compatible with Minkowski space-time: time as a dimensional representation applies to the universe: the universe is not in time. Likewise, space is in the universe: the universe is not in a region of space. Time (the dimension) takes space, and space (the dimensions) take time, and space-time is independent and unaffected by the absence of a physical instant and physical progression of time. To the contrary, as many readers will be aware, Minkowski space-time also illustrates time to be a derivative notion, not actually 'flowing' as our subjective conscious perceptions often seem to suggest.

## 3. Time and Classical and Quantum Mechanics: Indeterminacy vs. Discontinuity

The absence of a precise static instant in time underlying a dynamical physical process means that a body (micro and macroscopic) in relative motion does not have a precisely determined relative position at any time, and that all physical magnitudes are not precisely determined at any time, although with the parameter and boundary of their respective position and magnitude being determinable up to the limits of possible measurement as stated by the quantum hypothesis[1], but with this indeterminacy in precise value not being a consequence of $\hbar$ and quantum uncertainty. The reason why can be demonstrated by employing Albert Einstein's famous 1905 train and the other theoretical device it is associated with, the thought experiment. An observer is watching a train traveling by containing a young Albert Einstein. At any given time as measured by a clock held by the observer, Einstein's train is in motion. If the observer measures the train to pass a precisely designated point on the track at 10.00 seconds, this value indicates the train passes this point during the measured time interval of 10.00 and 10.00999…seconds. As Einstein's train is in motion at all measured times, regardless of how great or small its velocity and how small the measured time interval (i.e. 10.0000000-10.0000000999...seconds), Einstein's train does not have a precisely determined relative position to the track at any time, because it is not stationary at any time while in motion, for to have a precisely determined relative position at any time, the train would also need to be stationary relative to the track at that time. Conversely, the train does not have a precisely determined relative position at an ethereal precise static instant in time, because there is not a precise static instant in time underlying the trains motion. If there were, Einstein's trains motion would not be possible.

As the time interval measurement is made smaller and more accurate, the corresponding position the train can be said to 'occupy' during that interval can also be made smaller and more accurate. Momentarily forgetting $L_p$, $T_p$ and time keeping restrictions, these measurements could hypothetically be made almost infinitesimally small, but the train does not have a precisely determined position at any time as it is in motion at all times, regardless of how small the time interval. For example, at 100km/hr, during the interval of $10^{-25}$ s Einstein's train traverses the distance of $2.7^{-21}$ cm. Thus, it is exactly due to the train not having a precisely determined relative position to the track at any time, whether during a time interval, however small, or at a precise static instant in time, that enables Einstein's train to be in motion. Moreover, this is not associated with the preciseness of the measurement, a question of re-normalizing infinitesimals or the result of quantum uncertainty, as the trains precise relative position is not to be gained by applying infinitely small measurements, nor is it smeared away by quantum considerations. It simply does not have one. There is a very significant and important difference.

If a photograph is taken (or any other method is employed) to provide a precise measurement of the trains relative position to the track, in this case it does appear to have a precisely determined relative position to the track in the picture, and although it may also be an extremely accurate measure of the time interval during which the train *passes* this position or a designated point on the track, the imposed time measurement itself is in a sense arbitrary (i.e. 0.000000001 second, 1 second, 1 hour etc), as it is impossible to provide a time at which the train is precisely in such a position, as it is not precisely in that or any other precise position at any time. If it were, Einstein's train would not, and could not be in motion.

On a microscopic scale, due to inherent molecular, atomic and subatomic motion and resulting kinetic energy, the particles that constitute the photograph, the train, the tracks, the light radiation that propagates from the train to the camera, as well as any measuring apparatus e.g. electron microscope, clock, yardstick etc, also do not have precisely determined relative position's at any time. Naturally, bodies at rest in a given inertial reference frame, which are not constituted by further smaller particles in relative motion, have a precisely defined relative position at all measured times. However, as this hypothetical special case is relevant to only indivisible and the most fundamental of particles, whose existence as independent 'massive' objects is presently discredited by quantum physics and the intrinsic 'smearing' effects of wave-particle duality and quantum entanglement, if consistent with these considerations, this special subatomic case would not appear to be applicable. Furthermore, and crucially, as we shall see shortly, because once granted indeterminacy in precise relative position of a body in relative motion, also subsequently means indeterminacy in *all* precise physical magnitudes, including gravity, this also applies to the very structure of space-time, the dynamic framework in which all inertial spatial and temporal judgments of relative position are based. As such, the previously mentioned possible special case, isn't actually one, and the very same applies.

Consequently, the absence of a precise static instant in time underlying a dynamical physical process and the resulting lack of a precisely determined relative position at all times, but with the parameter and boundary of the position "occupied" being determinable up to the limits of the time interval and the corresponding position measurement, similarly applies to all physical magnitudes and values at all times: if displacement (relative position) *d* of a body in relative motion is not precisely determined at time *t*, neither is a velocity $v=\Delta d/\Delta t$, so neither is momentum $p=mv$, acceleration $a=\Delta v/\Delta t$, *g*, all rotational and angular kinematic magnitudes e.g. angular momentum, $L=Iw$ etc, wave speed *c*, frequency *f*, wavelength λ, period *T*, kinetic energy $E_k=1/2mv^2$, Schrodinger's wave equation $(-\hbar^2/2m \nabla^2 +v)\psi=i\hbar\, \partial\psi/\partial t$, current $I=Qt$, and so, charge $Q=I/t$, voltage $V=E/Q$ etc, time dilation $t=\gamma t_o$, length contraction $L=L_o/\gamma$, relativistic mass and relativistic momentum $p=\gamma mv$. Likewise, if *p* is not precisely determined, neither is de Broglie's matter wave $\lambda=h/p$. If $E_k$, relativistic mass and *c* are not precisely determined, according to $E=mc^2$, neither is rest mass and energy. If *f* is not precisely determined, neither is radiation energy per quantum $E=hf$. If *m, a* and *r* are not precisely determined, neither is Newton's universal gravitation $F=G$ ($M_1 m_2/r^2$), force $F=ma$, so neither is pressure $p=F/A$, electric field strength $E=F/Q$, impulse $F\Delta t=m\Delta v$, torque $T=Fx$, work $W=Fs$, $Fd$, so neither is power $P=W/t$, $E/t$, $Fv$, $VI$ etc. If $m$, $g$ and relative position $h$ are not precisely determined, neither is gravitational potential energy $E_p=mgh$, and in conjunction with $f$, according to Einstein's general theory of relativity, neither is a precise interval of time as indicated by a clock under the influence of gravity, relative to another clock. Moreover, if $v$, $f$, $E$ and $m$ are not precisely determined, neither is any physical magnitude, and as this includes gravity, this also applies to the very structure of space-time. Crucially however, this universal indeterminacy in *precise* physical magnitude is not a consequence of $\hbar$ and quantum uncertainty.

### 3.1 Newton and Zeno of Elea's Motion and Infinity Paradoxes

The only situation in which a physical magnitude would be precisely determined was if there *were* a precise static instant in time underlying a dynamical physical process and as a consequence a physical system were frozen static at that instant. In such a system an indivisible mathematical time value, e.g. 2s, would correctly represent a precise static instant in time, rather than an interval in time (as it is generally assumed to in the context of calculus, and traceable back to the likes of Galileo, and more specifically, Newton, thus guaranteeing absolute preciseness in theoretical calculations before the fact i.e. $\Delta d/\Delta t=v$). Fortunately this is not the case, as this static frame would include the entire universe. Moreover, the universe's initial existence and progression through time would not be possible. Thankfully, it seems nature has wisely traded certainty for continuity. [2]

Another way to look at this is if a physical value were precisely determined at a precise instant in time, it could never change, as it would firstly have to proceed to another precise value. But before it could do this, it would firstly have to proceed to half of that value. But before it could do this, it would have to proceed to half of that value again, and so on, and so on, to infinitum. Thus, in this manner it can be demonstrated that if a physical value were precisely determined, it could never change. There is a necessary trade off of between certainty at a time, for continuity through time. Please note that the explanation provided here and previously throughout this paper is also the correct solution to the motion and infinity paradoxes the Dichotomy, Achilles and the Tortoise, the Arrow, and their other more modern variations, originally conceived by the Greek mathematician, Zeno of Elea.[10] That is, they all have the same general solution through such reasoning as has been discussed here, and are not distinct and different problems requiring different and distinct proposed solutions as has historically been assumed. [3]

### 3.2 A Consistent Classical and Quantum Mechanical Description

In relation to quantum mechanics then, this conclusion illustrates that the relationship between $\hbar$, wave-particle duality, quantum entanglement etc, and the constant indeterminacy in all precise physical values due to the absence of a precise static instant in time, although indistinguishable for all practical purposes when quantifying the overall state of a physical system at a microscopic level, are in fact separate and distinct variables, playing quite separate and distinct roles. As such, a revised and seemingly more appropriate description is: all physical magnitudes are not ever precisely determined due to the absence of a precise static instant in time underlying a dynamically physical process, although with the parameter and boundary of the their respective value being determinable up to the limits of possible measurement as stated by the quantum hypothesis and Heisenberg's uncertainty principle,[1] but with this indeterminacy in precise magnitude not being a consequence of $\hbar$ and quantum uncertainty. Following with the introduction of uncertainty and/or statistics in quantum values due to $\hbar$, wave-particle duality, quantum entanglement etc. This illustrates that in relation to indeterminacy in precise physical magnitude, there is not a distinction between the large and macroscopic and the microscopic realm of quantum mechanics, in the sense that both, the micro and macroscopic, are *directly* subject to inherent indeterminacy, rather than just a case of the former underlying and contributing to the latter. In this regard, they are both actually a part of the same parcel, being inextricably linked.

---

[2] Please note that there is obviously no fault in the actual mathematics here, but rather in the historical assumption underlying them regarding determined physical magnitudes at a time and/or instant.

[3] For a detailed explanation of Zeno's paradoxes and their resolution, please see, Lynds, P. Zeno's Paradoxes: A Timely Solution, http://philsci-archive.pitt.edu/archive/00001197/

I would suggest that there is possibly much more to be gleaned from the connection between quantum physics and the inherent need for physical continuity, and even go as far to speculate that the dependent relationship may be the underlying explanation for quantum jumping and with static indivisible mathematical time values directly related to the process of quantum collapse. Time will tell.[4]

## 4. Time and Quantum Cosmology

Detailed calculations have been completed in the theoretical field of quantum cosmology in an attempt to elucidate how time may have "emerged" and "congealed" out of the 'quantum foam' and highly contorted space-time geometry's and chaotic conditions preceding Planck scale $(Gh/c^3)^{1/2}$ just after the big bang (new inflationary model).[2-7] More specifically, it has been tentatively hypothesized that it would require particularly special initial quantum configurations for the "crystallization" of time and the emergence of macroscopic (non-quantum) phenomena to be possible.[2-7] This conclusion however, illustrates that temporality wouldn't need to "emerge" at all, but would be present and naturally inherent in practically all initial quantum states and configurations, rather than a specific few, or special one, and regardless of how microscopic the scale.

As soon as there is any magnitude of space (as a property of mass-energy), you naturally get the time dimension by default. If there is no mass-energy, there is no space-time. Because the reason continuity is possible is due to there not being a physical instant and physical progression of time, it is not necessary for time to "emerge" in the first instance. The more appropriate question remains: how can mass-energy, and as such, space-time emerge?, simultaneously bringing continuity with it due to the absence of a physical instant and physical progression of time i.e. temporality or continuity would only be required to emerge from possible initial quantum configurations, states or histories in which time *were* a physical quantity.

### 4.1 Imaginary Time

This conclusion is also not consistent with the cosmological proposal of "Imaginary Time" and "no boundary condition",[2, 3, 5-7] both, as a consequence of the above consideration, and secondly, because it is the relative order of events that is relevant, not the direction of time itself. It is not possible to assert using a model of the universe that includes a description of the sum over histories or path integrals of the actual structure of space-time, that time goes in any direction, let alone at 90 degrees to real time or linear time and takes on some of the properties, or is identical to that of spatial dimensions at approximately Planck scale $(Gh/c^3)^{1/2} \sim 10^{-33}$ cm, $10^{-43}$ s, while still being bounded by the big bang (or possible big crunch, in a now seemingly obsolete closed universe) singularities in real or linear time, but having no boundaries in imaginary time. Neither real nor imaginary time exist in a consistent physical description, as time does not go in any direction.

It is the relative order of events that is relevant, not the direction of time itself. The order of a sequence of events can take place in either one order relative to its reverse order, or in the reverse order, relative to the first. It is not possible for the order of a sequence of events to be imaginary in the mathematical sense as it is logically contradictory and meaningless to describe the order of a sequence of events as being at right angles relative to that of another sequence of events. The opposite to this could be posited for the

---

[4] Considerations to do with time discontinuity presented in this paper also have relation to similar investigations performed in the framework of p-adic mathematical physics. By using p-adic numbers, instead of continuous real numbers, we can get models (with p-adic noncontinuous time) that have both classical and quantum features. Please refer to:

A. Y. Khrennikov, *Non-Archimedean analysis:quantum paradoxes, dynamical systems and biological models*. Kluwer Acad. Publishers, Dordreht, (1997). Especially Chapter 4. and pg. 164-166 on Heisenberg uncertainty relation.

S. Albeverio and A. Y. Khrennikov, *Representation of the Weyl group in spaces of square integrable functions with respect to p-adic valued Gaussian distributions*. J. of Phys. A, v. 29, pg. 5515-5527, (1996).

A. Y. Khrennikov, *Ultrametric Hilbert space representation of quantum mechanics with a finite exactness*. Found. of Physics, 26, No. 8, 1033-1054, (1996).

relative spatial direction of events, but events take place at right angles relative to others on a regular basis, and this has nothing to do with their direction, or the direction of time becoming imaginary.

The fact that imaginary numbers appear when computing space-time intervals and path integrals does not facilitate that when multiplied by *i*, that time intervals become basically identical to dimensions of space. Imaginary numbers show up in space-time intervals when space and time separations are combined at near the speed of light, and spatial separations are small relative to time intervals. What this illustrates is that although space and time are interwoven in Minkowski space-time, and time is the fourth dimension, time is not a spatial dimension: time is always time, and space is always space, as those *i*'s keep showing us. There is always a difference. If there is any degree of space, regardless of how microscopic, there would appear to be inherent continuity i.e. interval in time.

## 5. Chronons

"Chronons", proposed theoretical particles or atoms of indivisible intervals of time,[2, 8] are also not compatible with a consistent physical description. There is not a need for quantum or atomic indivisible time operators to stitch microscopic events together to facilitate physical continuity, as this overall conclusion illustrates that the very reason events are continuous in the first instance is due to there *not* being an atom or quantum of time. Simply, if there were, physical continuity, motion and time (relative interval) would not be possible.

## 6. Conclusion

In summary, it was shown there is a necessary trade off of all precisely determined physical magnitudes and values at a time, for their continuity through time, although with the parameter and boundary of their respective magnitude and value being determinable up to the limits of possible measurement as described by the quantum hypothesis,[1] but with this indeterminacy in precise value not being a consequence of $\hbar$ and quantum uncertainty. This illustrated that in relation to indeterminacy in precise physical magnitude, the macro and microscopic are inextricably linked, rather than being a variable only directly associated with the quantum world. The explanation provided was also shown to be the correct solution to the motion and infinity paradoxes, excluding the Stadium, originally conceived by the ancient Greek mathematician, Zeno of Elea.[9] It is not necessary for time to "emerge" from the "quantum foam" present just after the big bang at approximately $(Gh/c^3)^{1/2}$ scale,[2-7] and the proposals of "Imaginary Time",[2, 3, 5-7] and "Chronons",[2, 8] have been shown to be incompatible with a consistent physical description, and would appear to be superseded on a theoretical basis.

Conversations and correspondence with, and encouraging words received from J. A. Wheeler of Princeton, A. Khrennikov of Sweden, J. J. C. Smart and H. Price of Australia, C. Grigson of New Zealand, A. P. French of MIT and B. Yigitoz of Canada are most gratefully acknowledged.